\documentstyle[bibnorm]{lamuphys}
\input epsfig.sty
\begin{document}
\title{On Wilson loops and $Q\bar Q$-potentials from the AdS/CFT relation 
at $T\geq 0$}
December 1998\\
\begin{flushright}
HU Berlin-EP-98/71\\
\end{flushright}
\mbox{ }  \hfill hepth@xxx/9812109
\vspace{5ex}
\Large
\begin {center}
\bf{On Wilson loops and $Q\bar Q$-potentials from the AdS/CFT relation 
at $T\geq 0$\footnote{Based on talks at the conferences ``32$^
{\mbox{\tiny nd}}$ International Symposium Ahrenshoop on the Theory of 
Elementary Particles'' 
Buckow, September 1-5, 1998 and\\ "Quantum Aspects of Gauge Theories, 
Supersymmetry and Unification", Corfu, 20-26 September 1998}}
\end {center}
\large
\vspace{1ex}
\begin{center}
H. Dorn, H.-J. Otto
\footnote{e-mail: dorn@physik.hu-berlin.de otto@physik.hu-berlin.de}
\end{center}
\normalsize
\it
\vspace{1ex}
\begin{center}
Humboldt--Universit\"at zu Berlin \\
Institut f\"ur Physik, Theorie der Elementarteilchen \\
Invalidenstra\ss e 110, D-10115 Berlin
\end{center}
\vspace{4ex}
\rm
\begin{center}
{\bf Abstract}
\end{center}
We give a short introduction to and a partial review of the work on the 
calculation of Wilson loops and $Q\bar Q$-potentials via the conjectured 
AdS/CFT duality. 
Included is a discussion of the relative weight of the stringy correction
to  the target space background versus the correction by the quantum 
fluctuations of the string world sheet.
\vfill
\newpage  
\author{Harald\,Dorn \and Hans-J\"org\,Otto}

\institute{Institut f. Physik, Humboldt-Universit\"at Berlin, Invalidenstr. 110,
D-10115 Berlin, Germany}

\maketitle

\begin{abstract}
We give a short introduction to and a partial review of the work on the 
calculation of Wilson loops and $Q\bar Q$-potentials via the conjectured AdS/CFT duality. 
Included is a discussion of the relative weight of the stringy correction
to  the target space background versus the correction by the quantum 
fluctuations of the string world sheet.
\end{abstract}
\section{Wilson loops in gauge theory}
Wilson loops  
$W[{\cal C} ]=\mbox{tr P}\exp (i\int _{{\cal C}}A_{\mu}dx^{\mu})$
play an extremely crucial role in various aspects of gauge field dynamics.
In the context of this talk we have in mind the attempts to encode gauge field
dynamics in equations for functional derivatives of $W$ \cite{PolMM} and in 
particular the evaluation of quark-antiquark potentials from the Wilson loop 
for rectangular closed contours $\cal C $.

The static potential $V(L)$ between external colour sources (heavy quarks) 
separated by a distance $L$ is related to the Wilson loop for a rectangular $L\times t$ contour in Euclidean pure
gauge theory by \cite{Wilson}
\begin{equation}
V(L)~=~-\lim _{t\rightarrow\infty}\frac{1}{t}\log \langle W[{\cal C}]\rangle
~.
\label{1}
\end{equation}
A linear confining potential corresponds to the famous area law.

In $T>0$ equilibrium thermodynamics, described by $D=3+1$ dimensional
Euclidean QFT with a periodic dimension of period $\frac{1}{T}$, the 
role of the Wilson loops is twofold. At first the expectation value
$\langle W[{\cal C}]\rangle $ for a closed contour wrapping the compactified
direction (Wilson line, Polyakov loop) is an order parameter for 
con\-fine\-ment/de\-con\-fine\-ment. Secondly, one can read off the 
$Q\bar Q $-potential from the
correlation function of two Wilson lines related to contours ${\cal C}_1 $ 
and ${\cal C}_2 $ separated by the distance $L$
\begin{equation}
V(L,T)~=~-\log \langle W[{\cal C}_1]W[{\cal C}_2]\rangle ~.
\label{1a}
\end{equation}
\section{AdS/CFT representation of Wilson loops at $T=0$}
The AdS/CFT duality conjecture \cite{Malda} in its most familiar case
states the equivalence of type IIB string theory on $AdS_5\times S^5$
in the presence of $N$ units of flux of the RR 5-form to ${\cal N}=4$
$SU(N)$ super Yang-Mills field theory in $M_4$. The couplings of both
sides are related by $\frac{1}{2\pi }(\frac{1}{g}+i\frac{\chi}{2\pi })=\frac{1}{g^2_{YM}}+i\frac{\Theta}{8\pi }$.

The main motivation for the conjecture comes from the equivalence of the isometry group of $AdS_5\times S^5$ and the ${\cal N}=4$ superconformal group in 
$M_4$ and the dynamics of open strings in the presence of $N$ copies of
D3-branes. To get some intuition for the calculation of Wilson loops via
this duality we now shortly comment on this string dynamics. The metric
associated to $N$ such copies at coinciding position is given by
\begin{eqnarray}
ds^2&=&f^{-\frac{1}{2}}~d\vec x _{\vert\vert}^2~+~f^{\frac{1}{2}}(dr^2+r^2d\Omega _5^2)~,\nonumber\\
f&=&1~+~\frac{4\pi g N \alpha ^{\prime 2}}{r^4}~,~~~~~~~~r~=~\vert \vec x_{\bot}\vert ~.
\label{2}
\end{eqnarray}
Enumerating the D3-branes by a Chan-Paton index the dynamics of open
strings with Chan-Paton charges at their ends yields an $U(N)$ gauge
theory on the branes which decouples from the bulk of 10-dimensional
target space in the limit $\alpha ^{\prime}\rightarrow 0 $. Separating
a subset of branes realizes the stringy version of the Higgs effect.
E.g. $U(N+1)\rightarrow U(N)\times U(1)$ is manufactured by separating one
brane from the remaining $N$ branes. A string stretching between the separated brane and the other ones mimics a quark in the unbroken $U(N)$. The distance
in $u~=~\frac{r}{\alpha ^{\prime}}$ plays the role of a mass scale.

The metric (\ref{2}) interpolates between flat 10-dim Minkowski space at
$r\rightarrow\infty$ and $AdS_5\times S^5$ in the near brane region $r\rightarrow 0$. Performing the decoupling limit $\alpha ^{\prime}\rightarrow 0 $
at fixed $u$ one tests the near brane region. The metric expressed in
$\vec x _{\vert\vert}, u$ and the $S^5$-variables becomes with
$R~=~(4\pi gN)^{\frac{1}{4}}~=~(2g_{\mbox{\tiny YM}}^2 N)^{\frac{1}{4}}$
\begin{equation}
ds^2~=~\alpha ^{\prime}~\left ( \frac{u^2}{R^2}d\vec x_{\vert\vert}^2~+~
\frac{R^2}{u^2}du^2~+~R^2~d\Omega _5 ^2\right )~=~\alpha ^{\prime}~G_{MN}dx^Mdx^N~.
\label{3}
\end{equation}
The precise mapping for correlation functions of local operators has been
given in ref.\cite{WittenGKP} and those for the nonlocal Wilson loops
in ref.\cite{MaldaRey}. The analogous recipe for Wilson loops is
\begin{equation}
\langle W[{\cal C}]\rangle ~=~Z[{\cal C}]~,
\label{4}
\end{equation}
where ${\cal C}$ is the contour for the Wilson loop in $M_4\times S^5$,
realized at the boundary of $AdS_5$ at $u\rightarrow\infty $ and $Z[{\cal C}]$
is the string partition function for world sheets approaching the given 
${\cal C}$ at $u\rightarrow\infty $.

The factor $\alpha ^{\prime -1}$ in
the Nambu-Goto action is cancelled by the factor $\alpha ^{\prime}$
in eq.(\ref{3}), i.e.
\begin{equation}
S_{\mbox{\tiny NG}}~=~\frac{1}{2\pi}\int d^2z\sqrt {\mbox{det}(G_{MN}\partial _{\mu}
x^M\partial _{\nu}x^N)}~.
\label{4a}
\end{equation}
The string partition function can be approximated by the AdS-area 
of the stationary surface of lowest order topology for small string
coupling $g$ and small curvature of the target space (i.e. large $R$).
We call this the classical approximation. On the YM side
of the duality this limit corresponds to the t'Hooft limit
$N\rightarrow\infty ,~g_{\mbox{\tiny YM}}\rightarrow 0$ at large t'Hooft coupling
$g_{\mbox{\tiny YM}}^2N$.

Even in this classical approximation the calculation of the Wilson loop for generic contours is a highly nontrivial task. Up to now results are available
only for circular loops \cite{GrossSB, BCFM} and in the limiting case
for rectangles needed to evaluate $Q\bar Q$ potentials via (\ref{1}).
In the last case one needs the leading $t$-behaviour of a $L\times t$
rectangle. This becomes translation invariant in the $t$ coordinate and
instead solving a partial differential equation one can restrict to 
an ordinary one. The potential between static $Q$ and $\bar Q$ separated by
a distance $L$ and for $Q$ and $\bar Q$ both having constant orientation in internal $S^5$, with relative angle $\Delta\Theta $, has been calculated in 
ref. \cite{MaldaRey}
\begin{equation}
V(L,\Delta\Theta )~=~-\frac{2}{\pi}~\frac{(2g_{\mbox{\tiny YM}}^2N)^{\frac{1}{2}}}{L}
(1-l^2)^{\frac{3}{2}}\left (\int _1^{\infty}\frac{dy}{y^2\sqrt{(y^2-1)(y^2+1-l^2)}}\right )^2~.
\label{5} 
\end{equation} 
The quantity $l$ is a monotonic function of $\Delta\Theta $ with $l(0)=0$
and $l(\pi )=1$.

The potential is Coulombic. Hence the AdS/CFT conjecture passed a further
consistency check, since the $\frac{1}{L}$ behaviour is dictated by the conformal
invariance of ${\cal N}=4$ SYM. The calculation involved a subtraction
of the energy stored in two strings stretching from $u=0$ to $u=\infty $. The regularised version of (\ref{5}) obtained by positioning ${\cal C}$ at $u=\Lambda $ has Coulomb behaviour for $L\cdot \Lambda \gg 1$. This is another 
manifestation of the IR/UV relation within the AdS/CFT correspondence 
\cite{SW}. 

Formula (\ref{5}), valid for large t'Hooft coupling, is nonperturbative from 
the SYM point of view. The SYM perturbative potential is Coulombic as well, 
but has a factor $g_{\mbox{\tiny YM}}^2N$ instead of $(g_{\mbox{\tiny YM}}^2N)^{\frac{1}{2}}$. 
Higher order corrections with respect to the AdS curvature should interpolate
between both regimes. It has been argued in ref.\cite{GrossSB} that the 
same situation concerning the coupling constant dependence appears for
the factors in front of singularities due to cusps of the Wilson loop
contour ${\cal C}$.

A last comment concerns the dependence on $\Delta\Theta $. For opposite
orientation of $Q$ and $\bar Q$ in $S^5$ ($\Delta\Theta =\pi $) the
static force vanishes. This is in agreement with a corresponding BPS argument
\cite{MaldaRey}.
\section{$T>0$ and attempts to make contact with QCD}
To describe the situation with non-zero temperature  the metric (\ref{3})
has to be replaced by the corresponding near brane limit of the metric
of a set of $N$ coinciding non-extremal D3-branes leading to 
\cite{Witten, Brand, Theisen}
\begin{equation}
ds^2~=~\alpha ^{\prime}~\left (\frac{u^2}{R^2}[h(u)dx_0^2+dx_i^2]~+~
\frac{R^2}{u^2h(u)}du^2~+~R^2~d\Omega _5^2\right )~,
\label{6}  
\end{equation}
with $h(u)=1-\frac{u_T^4}{u^4}, ~~u_T=\pi R^2 T, ~~x_0$ periodic with period
$\frac{1}{T}$, $i=1,2,3$ and  $R$ as before. In these papers the stationary string world sheet
is constructed for two qualitatively different situations.

In the so-called time-like case the boundary at $u=\infty $ is given by
two lines at constant $x_i$, separated by a distance $L$ and wrapping the
compact dimension $x_0$. Under the AdS/CFT duality it is mapped to
the correlation function of two Wilson lines and via (\ref{1a})
to the static $Q\bar Q$-potential in 3-dim space in a heat bath at temperature
$T$.

In the second case, called space-like, the boundary at $u=\infty $ is a rectangle extending in the space dimensions $x_i$ only. It becomes of particular interest in connection with the proposal \cite{Witten} to use the breaking of
SUSY by the periodic/antiperiodic boundary conditions to reach nonsupersymmetric gauge theory for $T\rightarrow\infty$, i.e. in the limit of decoupling
compactified dimension. The resulting ``QCD'' lives in $(2+1)$ space-time
dimensions at $zero$ temperature. 

For the time-like case one gets a $Q\bar Q$-potential \cite{Brand, Theisen}
which behaves like $\frac{1}{L}$ at small $L$ and can be estimated
numerically for generic $L$. There appears a critical distance
$L_{\mbox{\tiny crit}}$ beyond which the $Q\bar Q$ force vanishes identically.
It has been argued that this total screening should be absent if all quantum 
corrections are taken into account \cite{GrossOo}.

In the space-like case the $Q\bar Q$-potential turns out to be linear in $L$ 
for large $L$. The factor $\sigma $ in front of $L$, tentatively called
QCD string tension, is equal to $\frac{\pi R^2T^2}{2}$ \cite{Brand, Theisen}.
Expressing $R$ in terms of $g_{\mbox{\tiny YM}}$ and $N$ and taking into account
the relation to the Yang-Mills coupling in the dimensionally reduced theory ($g_{\mbox{\tiny YM},3}^2=Tg_{\mbox{\tiny YM}}^2$) one gets
\begin{equation}
\sigma ~=~\frac{\pi R^2T^2}{2}~=~\frac{\pi}{\sqrt{2}}g_{\mbox{\tiny YM},3}~N^{\frac{1}{2}}
~T^{\frac{3}{2}}~.
\label{7}
\end{equation} 
The whole setup has been generalised to D-branes of arbitrary dimension.
In particular, to end up with $\mbox{``QCD''}_4$ one has to start with
a set of D4-branes. The tension then turns out to be $\frac{8}{27\pi ^2}
g_{\mbox{\tiny YM},4}^2NT^2$ \cite{Brand, GrossOo}.

Only for $T\rightarrow\infty $, where the compactified dimension decouples, there is a chance to make contact with renormalised QCD. $T\rightarrow\infty $
and finite tension $\sigma $ requires \\
$R\rightarrow 0$. In this limit one
obviously runs outside the range of applicability of the classical approximation. Reaching QCD requires substantial progress in calculating the higher
order $\frac{1}{R}$-corrections. Another problem for reaching realistic QCD
is connected with the possibility of a phase transition in varying $N$
\cite{Witten, GrossOo, Li}.

The background (\ref{6}) has been used for a calculation of glueball 
masses \cite{Csaki, CsakiMin}, too. There it turns out that the mass of the KK excitations due
to the compactification are of the order of the glueball masses. This obstacle might be overcome by using modified backgrounds containing a second scale
\cite{Russo} as it is done in ref.\cite{CsakiMin}.
The Wilson loop calculation in such multi-scale backgrounds has been
included in ref.\cite{Sonn2}. They also yield linear confining potentials.

A last point, which has been raised \cite{Olesen} in connection with a comparison to QCD, is the absence of any universal L\"uscher-type $\frac{1}{L}$ 
term \cite{Lue} 
in the ``QCD'' potentials derived so far. As argued in
\cite{KalTs, Sonn} such a term should be connected with the quantum fluctuations of the string world sheet, similar to the original situation in \cite{Lue}.
\section{Corrections to the classical approximation}
Leaving aside the contributions from higher genus string world surfaces,
there are for fixed (lowest order) genus two sources of corrections to
the classical approximation. 

At first the target space background which was constructed out of a D3-brane
solution of type IIB supergravity should be replaced by the similar 
construction based on D3-brane solutions of the stringy effective action 
to all orders in  $\alpha ^{\prime}$. For $T=0$ there are no such corrections
\cite{Green}. For $T>0$ they have been studied in ref. \cite{GKT}.
Their nextleading term has a factor $(\alpha ^{\prime})^3 L_{\mbox{\tiny throat}}^{-6}$
relative to the leading term. Expressing $L_{\mbox{\tiny throat}}$ by our parameter $R$,
$\alpha ^{\prime}$ cancels and the relative weight is $R^{-6}$, up to a 
numerical factor. 

The second source of corrections are quantum fluctuations of the string
world sheet. Their dependence on $R$ is most easily estimated by changing
variables in (\ref{6}) by $u~=~\frac{R^2}{v}$, resulting in
\begin{equation}
ds^2~=~\alpha ^{\prime}~R^2~\left (\frac{1-\pi ^4T^4v^4}{v^2}dx_0^2~+~\frac{1}{v^2}dx_i^2~+~\frac{1}{v^2(1-\pi ^4T^4v^4)}dv^2~+~d\Omega _5^2\right )~.
\label{8}
\end{equation}
The boundary of the string world sheet now is at $v=0$, the other coordinates 
describing the contour ${\cal C}$ are not changed. For $T=0$ this metric is a 
conformal flat version
of $AdS_5\times S^5$. Now $R^2$ appears as an overall factor in front of the Nambu-Goto
action $S_{\mbox{\tiny NG}}$ for the metric given by (\ref{8}) without
the factor $\alpha ^{\prime}~R^2$. For $R\rightarrow \infty $ the expansion of the string partition function $Z[{\cal C}]$ has the form ($~R^{-\infty}$ 
put into the normalization)
\begin{equation}
Z[{\cal C}]~=~e^{-R^2~S_{\mbox{\tiny NG}}[x_{\mbox{\tiny class}}]}~\left ( \det\frac{\delta^2S_{\mbox{\tiny NG}}}
{\delta x_{\mbox{\tiny class}}\delta x_{\mbox{\tiny class}}}\right )^{-\frac{1}{2}}~(1~+~O(R^{-\frac{1}{2}}))~.
\label{9}
\end{equation} 
The relative strength of the nextleading (determinant) contribution in comparison to the leading classical contribution is up to numerical factors
$R^{-2}$. Therefore, assuming the absence of numerical peculiarities, the
contribution from quantum fluctuations should dominate the corrections due to stringy corrections of the classical target space background.

Although more important, up to now the quantum fluctuations have not been calculated due to technical difficulties. On the other side the extension of the
classical calculations of refs.\cite{MaldaRey, Brand, Theisen} is straightforward and has been done in our paper \cite{DO}. We used the metric ($h(u)=1-
\frac{u_T^4}{u^4}$) 
\begin{equation}
ds^2~=~\alpha ^{\prime}\left (\frac{u^2h(u)}{R^2}~e^{\gamma A}dx_0^2+\frac{u^2}{R^2}~e^{\gamma C}dx_i^2+\frac{R^2}{u^2h(u)}~e^{\gamma B}du^2
+R^2~e^{\gamma D}d\Omega _5^2\right )~,
\label{10}
\end{equation} 
with $u_T=\pi R^2T(1+15\gamma )^{-1}$ and $A,B,C,D$ polynomials in 
$\frac{u_T}{u}$ with known
coefficients. We took $\gamma =0$ to describe the uncorrected classical
approximation $or$ $\gamma ~=~\frac{1}{8}\zeta (3)R^{-6}$ to include the
first nontrivial stringy correction \cite{GKT}.

Our main focus was both the inclusion of the $\Delta\Theta $ internal space
dependence into the $T>0$ calculation and studying the effect of switching on the nextleading background correction.

In the timelike case we found a critical line in the $(L,\Delta\Theta )$-plane
beyond which the $Q\bar Q$-force is screened. The absence of a force for
$all$ distances at antipodal internal orientation $\Delta\Theta =\pi $,
known from $T=0$, holds to continue also for $T>0$. This is a nontrivial
fact, since due to broken SUSY the absence of a force is no longer
guaranteed by a BPS argument. The critical line is driven to larger values of $L$ if the correction is switched on ($\gamma >0$), see fig.1. This is a movement in the right direction in the sense of ref. \cite{GrossOo}.

In the spacelike case we found for $L\rightarrow\infty $
\begin{equation}
V(\Delta\Theta ,L)~=~\frac{\pi R^2T^2}{2}(1-\frac{265}{8}\gamma )\cdot L~+~
\frac{1}{4\pi}R^2(\Delta\Theta )^2(1+\frac{15}{8}\gamma )\cdot \frac{1}{L}~+~O(1
)~.
\label{11}
\end{equation}
The tension of the  ``QCD''-string is independent of $\Delta\Theta $.
The nextleading term in the $L$-asymptotics is $\propto \frac{1}{L}$.
Due to its dependence on the YM-coupling via $R$ and its sign it is no
L\"uscher-type \cite{Lue} term. Furthermore, there is in QCD no place for the 
$\Delta\Theta$ dependence, which is a remnant of ${\cal N}=4$ SUSY.
Hence it is gratifying that the term under discussion drops out
in the limit $T\rightarrow\infty ,~R\rightarrow 0$ which has been identified 
before as the limit necessary to make contact with renormalised QCD.
\begin{figure}
\begin{center}
\mbox{\epsfig{file=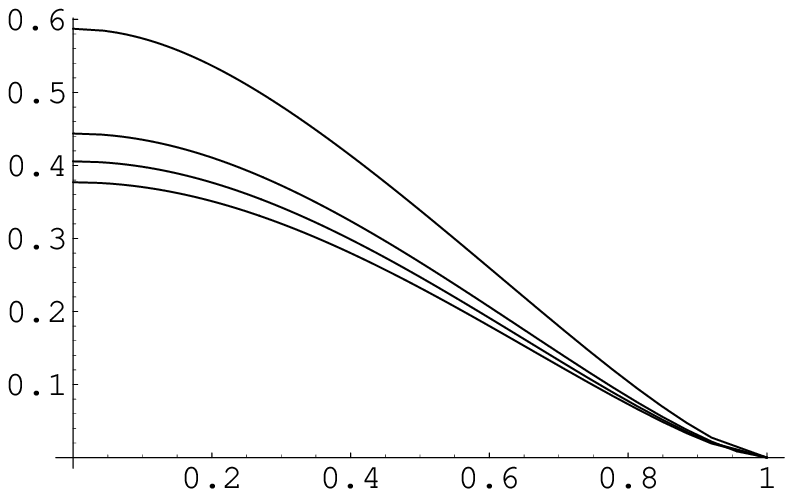 , width=60mm}}
\end{center}
\noindent {\bf Fig.1}\ \ {\it $\frac{\pi}{2}T\cdot L_{\mbox{\tiny crit}}$
as a function of $\frac{1}{\pi}\cdot\Delta\Theta $. 
The lowest line is for
$\gamma =0$, the others for\\
\hspace*{61mm}$\gamma = 0.01$, $0.02$ and $0.05$, respectively.}
\end{figure}

%
%

\end{document}